\documentclass[letterpaper,12pt]{article}   
\usepackage{osajnl2} 
\usepackage{graphicx}

\begin{document}

\title{Observation of a stronger-than-adiabatic change of light trapped in an ultrafast switched GaAs-AlAs microcavity}

\author{Philip J. Harding}
\affiliation{Complex Photonic Systems (COPS), MESA+ Institute for Nanotechnology, University of Twente, The Netherlands}
\author{Huib J. Bakker}
\affiliation{FOM Institute for Atomic and Molecular Physics (AMOLF), Science Park 104, 1098 XG Amsterdam,
The Netherlands}
\author{Alex Hartsuiker}
\affiliation{FOM Institute for Atomic and Molecular Physics (AMOLF), Science Park 104, 1098 XG Amsterdam,
The Netherlands}
\author{Julien Claudon}
\affiliation{CEA/INAC/SP2M, Nanophysics and Semiconductor Laboratory, 17 rue des Martyrs, 38054 Grenoble C\'edex, France}
\author{Allard P. Mosk}
\affiliation{Complex Photonic Systems (COPS), MESA+ Institute for Nanotechnology, University of Twente, The Netherlands}
\author{Jean-Michel G\'erard}
\affiliation{CEA/INAC/SP2M, Nanophysics and Semiconductor Laboratory, 17 rue des Martyrs, 38054 Grenoble C\'edex, France}
\author{Willem L. Vos}
\affiliation{Complex Photonic Systems (COPS), MESA+ Institute for Nanotechnology, University of Twente, The Netherlands}

\date{Prepared 13 December 2011}

\begin{abstract}
We study the time-resolved reflectivity spectrum of a switched
planar GaAs-AlAs microcavity. Between 5 and 40 ps after the switching (pump) pulse we observe a strong excess probe reflectivity and a change of the frequency of light trapped in the cavity up to 5 linewidths away from the cavity resonance.
This frequency change does not adiabatically follow the fast-changing cavity resonance. The frequency change is attributed to an accumulated phase change due to the time-dependent refractive index. An analytical model predicts dynamics in qualitative agreement with the experiments, and points to crucial parameters that control future applications.
\end{abstract}


\maketitle

\section{Introduction}
There is a fast growing interest to control and manipulate information encoded as photons as this promises much broader information bandwidths than with electronic integrated circuits~\cite{Noda:03}. Therefore it is important to realize time-dependent control of photonic systems, in other words, photonic switching. A crucial step is to have control and signal photons interact, such that the signal is phase modulated and its frequency changed to a different value. As the interaction between photons in free space is exceedingly weak~\cite{Breit:46}, one resorts to well-known methods of nonlinear optics to control light with light~\cite{Franken:61, Boyd:08}. Recently the switching of photonic bandgap crystals and nanophotonic cavities has been advocated in nonlinear optics~\cite{Johnson:02, Reed:03, Yanik:04, Notomi:06, Gaburro:06, Preble:07}. These devices permit a judicious tweaking of optical dispersion, offer greatly enhanced field strength in tiny volumes, and have small sizes that make them amenable to large-scale integration.

In nanophotonic nonlinear optics~\cite{Preble:07,McCutcheon:07,Tanabe:09}, a cavity is charged by a probe pulse \emph{before} a pump pulse arrives. The light in the cavity then adiabatically follows the cavity resonance and exits at the frequency where the cavity has quickly been tuned to. In this paper, we explore conditions where light does not adiabatically follow a single-resonance cavity, \emph{after} an earlier pump pulse. As a result, the frequency of probe light changes to a value different from the cavity resonance. To the best of our knowledge, such nanophotonic tuning of the light's frequency has not been observed before. 

\section{Sample and experimental setup}
\begin{figure}
\begin{center}
\includegraphics{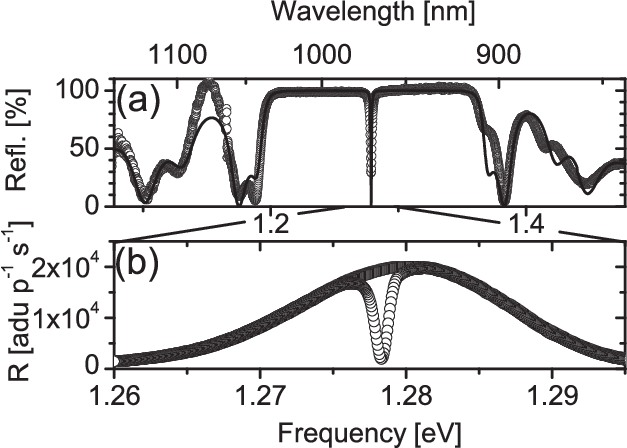}
\caption{(a) Broad-band reflectivity spectrum measured with a white-light source.
The reflectivity is calibrated by normalizing to the spectrum of a gold mirror.
The trough at 1.278 eV is the cavity resonance, the broad intense peak between 1.192 and 1.376 eV is the photonic stop band of the Bragg mirrors.
(b) High-resolution reflectance spectrum measured with the probe laser showing the cavity resonance at 1.278 eV. The width of the cavity resonance is slightly broadened due to wavevector spreading and interface roughness caused by imperfect growth (1.03 meV) compared to the homogeneous linewidth (0.85 meV).}
\label{fig:cwspectrum}
\end{center}
\end{figure}

We have studied a planar microcavity consisting of a GaAs $\lambda$-layer with a thickness $d = 275.1 \pm 0.1$ nm that is sandwiched between two Bragg mirrors made of 12 and 16 pairs of $\lambda/4$ thick layers of nominally pure GaAs and AlAs, see Refs.~\cite{Harding:07, HardingLineshape:09, Hartsuiker:09}. A normal-incidence reflectivity spectrum reveals the single cavity resonance at $E_{\rm{cav}} = 1.278$ eV ($\lambda = 970.2$ nm) centred in the $14.3 \%$ wide stopband of the Bragg mirrors, see Fig.~\ref{fig:cwspectrum}(a). The cavity resonance in Fig. \ref{fig:cwspectrum}(a) has an inhomogeneous linewidth of 1.03 meV, as a result of a finite numerical aperture, and of a slight spatial variation of the resonance frequency. In separate time-resolved experiments, we have determined the average trapping time for photons in the cavity to be $\tau_{\rm{cav}} = 780 \pm 50$ fs, corresponding to a quality factor $Q = 1500 \pm 100$, and a homogeneous cavity linewidth $\Delta E_{\rm{cav}} = 0.85$ meV~\cite{Hartsuiker:09}.

Time-resolved pump-probe spectroscopy was performed on the cavity in reflection geometry~\cite{Euser:09}. Two independently tunable optical parametric amplifiers generate pump and probe pulses with $140 \pm 10$ fs duration. The pump pulses are incident obliquely to avoid scattering into the detection path (Fig.~\ref{fig:PosDeltat}). The pump frequency was not resonant with the cavity and tuned to 0.72 eV ($\lambda = 1720$ nm) to switch the refractive index in all GaAs layers (bandgap 1.42 eV) by exciting free carriers by two-photon absorption~\cite{Euser:05}; the AlAs is hardly excited as its bandgap is much larger (2.2 eV). Also, by setting the pump to the two-photon regime, we evade non-degenerate two-photon absorption at coincidence \cite{HardingNGTP:09}. Our observations of the change in frequency were reproduced in runs with pump frequencies below 0.62 eV in the three-photon regime of GaAs. To ensure that the microcavity is spatially homogeneously switched~\cite{Euser:05}, we focussed the normal-incident probe beam to a much smaller spot (28 $\mu$m Gaussian diameter) than the pump spot (113 $\mu$m.) The probe fluence of $1.0 \pm 0.3$ mJ.cm$^{-2}$ energy per pulse was kept well below the pump fluence of $30 \pm 3$ mJ.cm$^{-2}$ to prevent inadvertent pumping by the probe pulses. The probe pulses have a relative spectral bandwidth of $2.7 \%$ (1.260 to 1.295 eV) that is much broader than the cavity's bandwidth of $1/Q = 0.07\%$ as shown in the unreferenced reflectance spectrum in Fig.~\ref{fig:cwspectrum}(b). Thus, we probe the dynamic cavity resonance by resolving the transient reflectivity with a high-resolution (0.12 meV) spectrometer at each pump-probe delay $\Delta \tau$. The stepsize in the delay in Figs.~\ref{fig:CmpMeasTheory} was $1$ ps between $-2$ and $+8$ ps, and $10$ ps henceforth. We used a PI/Acton sp2558 monochromator and a 1024 channels OMA-V InGaAs diode array. Due to inadvertent misalignment a slant occurred in the raw spectra that was corrected with a straight line and assuming constant $100 \%$ reflectivity in the stopband as in Fig.~\ref{fig:cwspectrum}. Spectra are averaged over $2\times 250$ probe pulses alternatingly reflected from a pumped and an unpumped cavity; the reproducibility of the unswitched resonance confirms that the cavity is unaffected by the experiment. The unswitched resonance also provides an \emph{in-situ} reference for the switched events. 

\section{Time-resolved transient reflectivity}
\begin{figure}
\begin{center}
\includegraphics[scale=0.5]{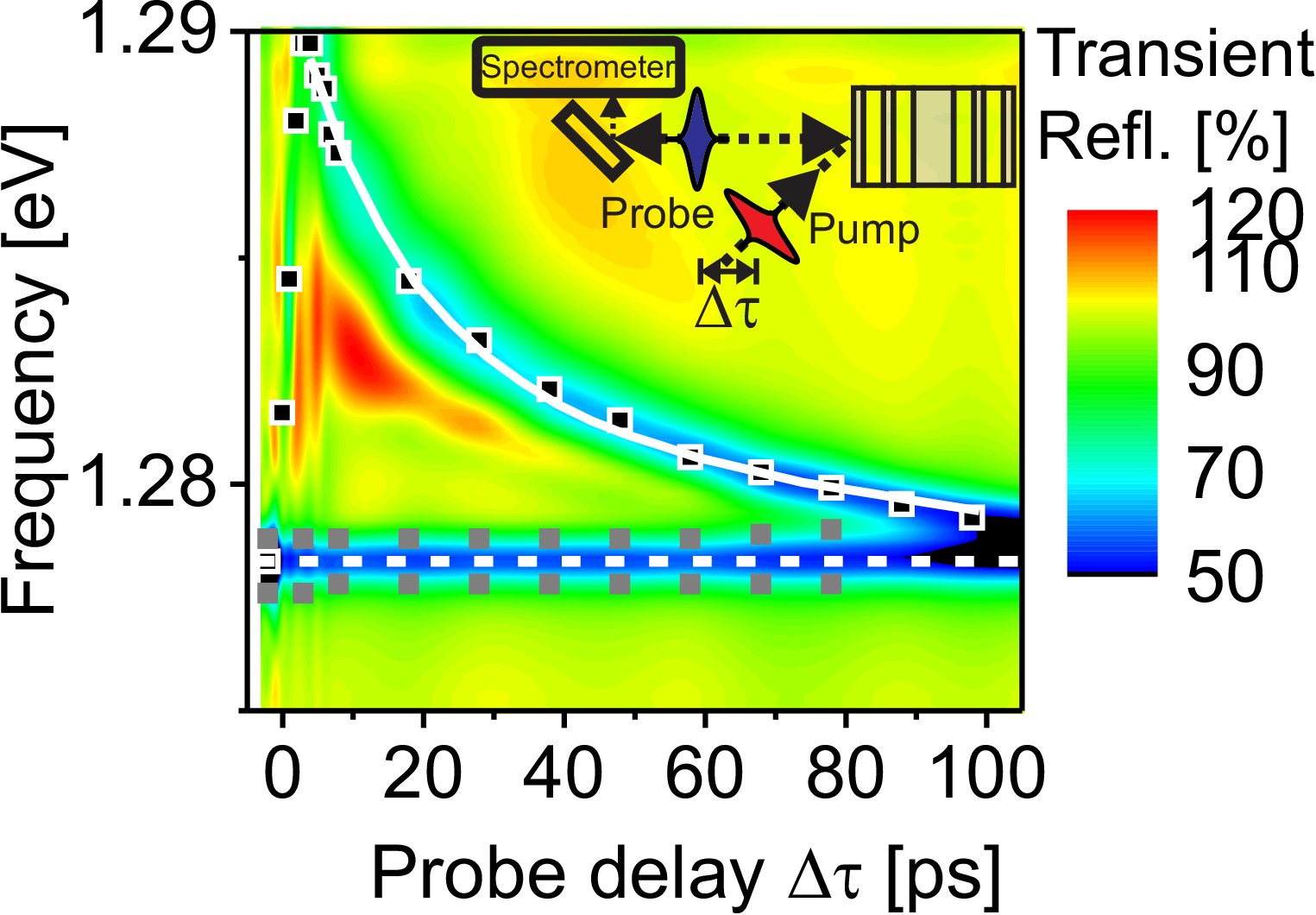}
\caption{(color) Transient probe reflectivity versus pump-probe delay and frequency. The spectra are averaged over both switched and unswitched events~\cite{note:signal}. The switched cavity resonance is shown by white squares. The solid white curve is a model of the switched cavity resonance using Drude free-carrier dispersion~\cite{HardingLineshape:09}.
The dashed white line is the unswitched cavity resonance. The faint modulation with delay (at 30, 60, 90 ps) is an artifact due to environmental changes during the scan. Inset: scheme of the setup with pump pulses entering obliquely and probe pulses reflected at normal incidence with delay $\Delta \tau$.}
\label{fig:PosDeltat}
\end{center}
\end{figure}

The cavity's time-dependent resonance is presented in a photon energy versus pump-probe time delay ($E, \Delta \tau$)-diagram shown in Fig.~\ref{fig:PosDeltat}. These data consist of many transient reflectivity spectra while stepping the pump-probe delay. The transient reflectivity is the time-integrated signal~\cite{Euser:09} that exits the cavity as a result of a short probe pulse incident at each $\Delta \tau$ (and should not be confused with the instantaneous reflectivity at $\Delta \tau$~\cite{note:signal}). The feature at 1.278 eV independent of time delay is the unswitched cavity resonance. When the cavity is optically switched with a short pump pulse, free carriers are excited leading to an ultrafast reduction of the refractive index. As a result, the cavity resonance frequency rapidly increases within 3 ps from $E_{\rm{cav}} = 1.278$ to 1.290 eV, while the cavity linewidth hardly changes~\cite{Harding:07}. This resonance shift corresponds to as much as 14.5 linewidths $\Delta E_{\rm{cav}}$, a signal of a strongly switched cavity. Subsequently, the carriers relax and the refractive index increases to its unswitched value in about 100 ps~\cite{Harding:07}.

\begin{figure}
\begin{center}
\includegraphics[scale=0.3]{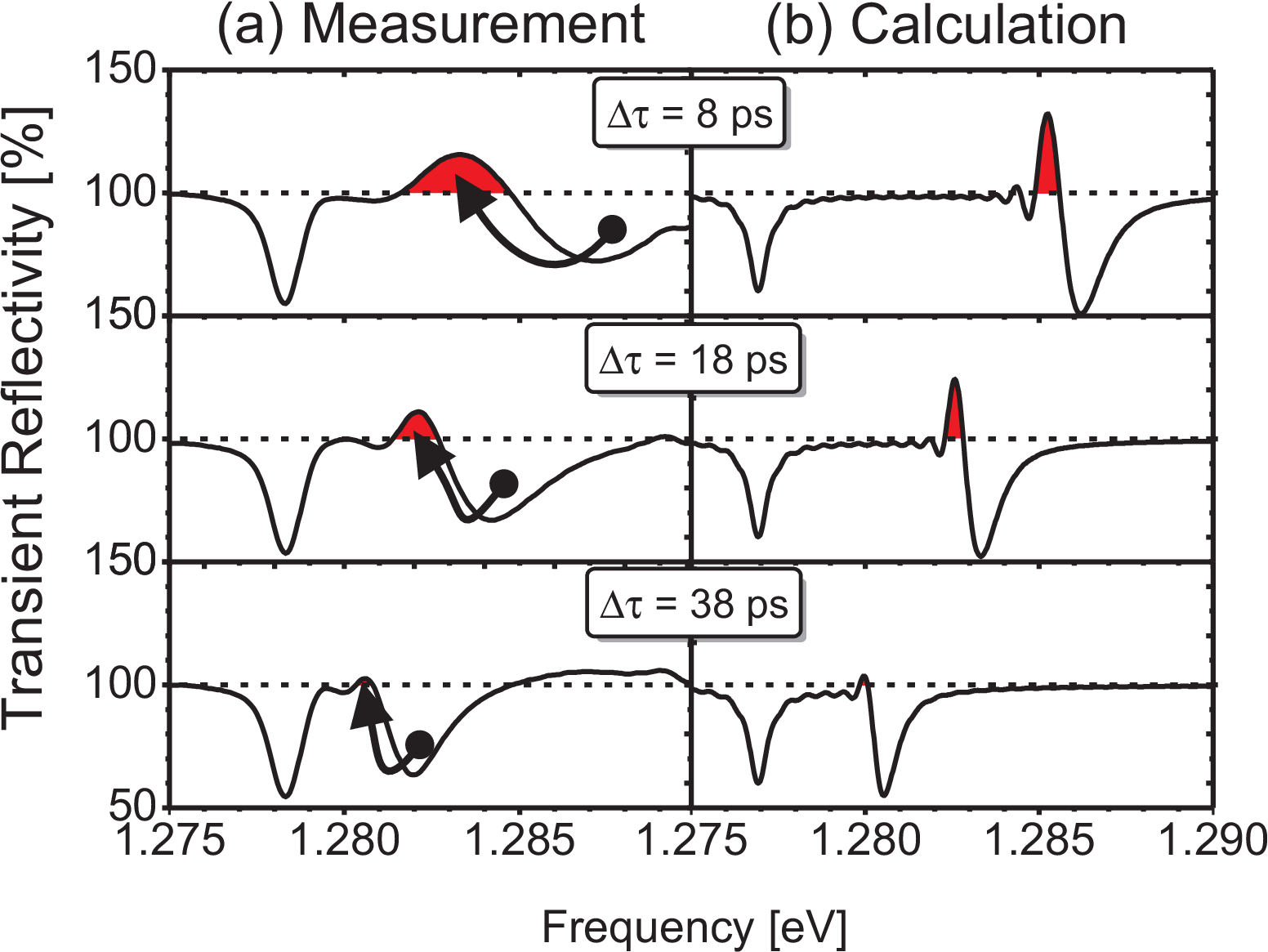}
\caption{(a) Transient probe reflectivity spectra measured at pump-probe delays $\Delta \tau=$ 8, 18, 38 ps (cross-sections from Fig.~\ref{fig:PosDeltat}).
Arrows indicate the frequency change of light from the cavity resonance to the excess transient reflectivity peaks (filled red). (b) Calculated transient reflectivity spectra at the same delays also reveal excess transient reflectivity peaks.}
\label{fig:CmpMeasTheory}
\end{center}
\end{figure}

To investigate the spectral features in more detail, we plot in Figure \ref{fig:CmpMeasTheory}(a) transient reflectivity spectra at select time delays. The trough at 1.278 eV in all spectra is the resonance of the unswitched cavity obtained from the events in absence of a pump. The trough of the switched cavity resonance ($E_{\rm{cav}} = 1.2876$ eV at 8 ps) is asymmetric due to the time-integrating nature of the experiment and the dynamics of the cavity resonance: photons exiting from the cavity are recorded while the resonance is shifting. Strikingly, the transient reflectivity exceeds $100\%$ to the red of the switched resonance, in particular $131 \%$ near 1.283 eV at 8 ps delay (and $122 \%$ at 18 ps delay and 1.282 eV)~\cite{note:signal}. The excess transient reflectivity corresponds to frequency shifted probe photons that enter the cavity at a delay $\Delta \tau = 8$ ps when the dynamic cavity resonance is still centred at 1.2876 eV. The frequency of the trapped probe photons has changed by as much as 4.6 meV or 5.4 linewidths to 1.283 eV when they exit at a later time $\Delta \tau' = \Delta \tau + \tau_{\rm{cav}}$.  

We make four salient observations: a.) The frequency of frequency-changed photons is {\em{substantially}} different from that of the cavity resonance at $\Delta \tau'$: During the trapping time $\tau_{\rm{cav}}$, the cavity shifts by $(dE_{\rm{cav}}/dt) \tau_{\rm{cav}}$, which is around 0.4 meV, and thus much smaller than the frequency change of the trapped light (4.6 meV at $\Delta \tau = 8$ ps). We therefore conclude that the frequency of the trapped photons does not adiabatically follow the cavity resonance. b.) Since the cavity is not resonant with the pump beam, the pump pulses have long vanished at $\Delta \tau = 8$ ps. Therefore, the change in the frequency of light is not caused by the temporal overlap with a pump pulse, in contrast to (nanophotonic) nonlinear optics~\cite{Boyd:08, Preble:07, McCutcheon:07, Tanabe:09}. c.) Since the frequency-shifting occurs long after the pump pulse, the quality factor $Q$ of the cavity does not degrade during the trapping time $\tau_{\rm{cav}}$. Quite the opposite, $Q$ actually increases slightly during $\tau_{\rm{cav}}$, in contrast to other switching schemes \cite{Preble:07, McCutcheon:07, Tanabe:09}. d.) The frequency-shifting of the trapped photons seems to correlate with the rate of change of the cavity resonance $dE_{\rm{cav}}/dt$ that characterizes the rate of change of the refractive index:  At short probe delays, the rate of change of the resonance is of the order of the characteristic ratio $\Delta E_{\rm{cav}}/dt$. At long probe delays $\Delta\tau > 60$ ps, where $dE_{\rm{cav}}/dt << \Delta E_{\rm{cav}}/\tau_{\rm{cav}}$, no frequency-shifted photons are observed. In summary, the observations for the frequency change of light in a strongly and ultrafast switched cavity differs fundamentally from previous frequency shifting as studied in, e.g., Refs. \cite{Boyd:08, Gaburro:06, Preble:07, McCutcheon:07, Tanabe:09}.

\section{Model for the frequency change of trapped light}
The frequency change of light in a strongly and ultrafast changing cavity can be calculated as follows without assumptions on adiabaticity in the following: Photons from the short probe pulse that enter the cavity are trapped for a while - on average the cavity trapping time $\tau_{\rm{cav}}$ - before exiting. The trapped light experiences a changing refractive index, which results in a phase shift $\Phi$ that changes the light frequency, whereas the cavity resonance shifts by a different amount. If the probe pulse enters the cavity at delay $\Delta \tau = t_{0}$, each angular frequency component $\omega=E/\hbar$ travels a distance $z$ (or correspondingly a time $nz/c$) through the cavity and acquires a phase shift 

\begin{equation}
\Phi(z) = E/(\hbar c) \int_{0}^{z} dz' (n(z') - n(0)),
\label{eq:phase}
\end{equation}

with $n(0)$ the refractive index at time $t_{0}$, and $n(z')$ the refractive index at time $t = t_{0}+z'n/c$. The refractive indices are determined by the pump-probe time delay $\Delta \tau$ and by the dynamics of the refractive index (here: dynamics of relaxing free carriers). The phase shift $\Phi(z)$ as a function of propagation length $z$ affects the time dependence of both the amplitude and phase profile of the exiting probe light ${\cal E}(t)$, equal to

\begin{equation}
{\cal E}(t) = \int_{0}^{\infty}dz' R(z') \int d\omega {\cal
E}_{0} (\omega) e^{i(\omega n z'/c -\omega t) + i\Phi(z')}.
\label{eq:et}
\end{equation}

Here frequency components are defined by the spectrum of the incident probe pulse ${\cal E}_{0} (\omega)$. The function $R(z')$ is the time-evolution kernel (Fourier transform of the S-matrix) that determines the range of cavity propagation lengths and hence the temporal profile of the light exiting from the cavity. Thus, light exiting at later times accumulates more phase shift $\Phi(z)$ since the interaction length is longer compared to light that exits early on. The accumulated phase is positive as the refractive index increases while the light ${\cal E}(t)$ exits from the cavity. Since the change in frequency is related to minus the time derivative of the phase ($\Delta E = -\hbar/e \cdot d\Phi(t)/dt$), the increased phase causes the frequency of the light to decrease.

The dynamics of the light field in a multilayer sample is described by a straightforward generalization of eq. \ref{eq:phase} that takes into account reflection and transmission coefficients of the 59 interfaces of the GaAs and AlAs layers in the sample. The refractive index of the GaAs layers are taken to be switched by Drude free-carrier dispersion to have the resonance frequency match the experiments (Fig.~\ref{fig:PosDeltat}). The time dependent refractive indices in the GaAs $\lambda$ layer and those in the GaAs mirrors have been calculated in Ref. \cite{HardingLineshape:09}. Equation \ref{eq:et} was solved numerically, where $R(z')$ was determined by solving Maxwell's equations for the 12 and 16 GaAs/AlAs Bragg mirrors with the time dependent refractive indices as above. The width of the cavity resonance determined by this exact solution does not account for interface roughness caused by imperfect growth and is therefore slightly narrower (0.74 meV) than the homogeneous linewidth which is gratifying as there are no free parameters. We verified that for $\Phi=0$ no frequency changes occur. While the accumulated phase per cavity roundtrip is small due to the small change in $n$ during one roundtrip, the phase accumulates substantially within $2Q$ roundtrips. Fig.~\ref{fig:CmpMeasTheory}(b) shows the power spectrum of the waveforms ${\cal E}(t)$ at the same delays as in Fig.~\ref{fig:CmpMeasTheory}(a). For better comparison with the experiment we average over switched and unswitched spectra. At $\Delta \tau = 8$ ps, we find the unswitched cavity resonance at 1.278 eV, and the switched cavity resonance at 1.287 eV. The switched cavity resonance exhibits a strong asymmetry caused by the time-integration of the shifting cavity resonance that we also observe in the experiment. The redshift of the trapped light explains the occurrence of transient reflectivity peaks $>$100\% at frequencies below the switched cavity resonance at 1.286 eV and 8 ps in Fig.~\ref{fig:CmpMeasTheory}(b). The experimental shifts are larger than in the model, which we tentatively attribute to beyond-Drude carrier dynamics in the experiments since the carrier density ($\sim 10^{19}$ cm$^{-3}$) approaches the range where scattering of electrons with other electrons, holes, or phonons become significant~\cite{Hugel:97}. The stronger-than-adiabatic response could possibly also be caused by an interaction of the weak probe pulses with the free carrier plasma. The qualitative phenomena of the spectra are however fully captured without such an assumption. Overall, we find a good qualitative agreement between the theoretical and experimental curves. 

\begin{figure}
\begin{center}
\includegraphics[scale=0.7]{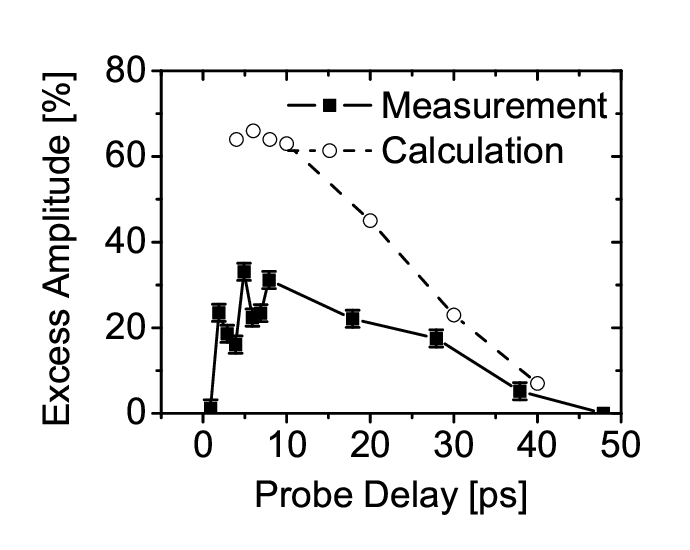}
\caption{Excess transient reflectivity versus pump-probe time delay $\Delta \tau$.
Squares are our experimental results~\cite{note:signal}, and the circles denote results from our analytical model.}
\label{fig:ExcAmp}
\end{center}
\end{figure}

Fig.~\ref{fig:ExcAmp} shows the excess reflectivity (or intensity of frequency changed light) versus delay. Both the model and the measurements reveal a maximum amplitude at small pump-probe delays of 5 to 10 ps. We propose that the excess reflectivity is large when the rate of change of the resonance $dE_{\rm{cav}}/dt$ is substantial compared to the cavity's characteristic ratio $(\Delta E_{\rm{cav}} / \tau_{\rm{cav}})$. The time dependence of the calculated excess reflectivity agrees well with the measured data, which indicates that the dynamics of the trapped light and of the refractive indices are correctly modeled. With increasing delay the rate of change of the refractive index decreases, hence the phase modulation $\Phi$ of the trapped probe light and the excess reflectivity decrease.

\section{Discussion}
Previously, ultrafast switching of a nanophotonic cavity has been discussed in the context of the adiabatic theorem in quantum mechanics~\cite{Bohm:89}. In quantum mechanics, adiabatic and non-adiabatic indicate the ability for a high-frequency degree of freedom to follow or not a change in a low-frequency degree of freedom, such as a 2-level spin in an external magnetic field. In the adiabatic case, the change of the field is sufficiently slow for the occupied quantum state to remain in the same state. In the non-adiabatic case, the change is so fast that the occupied state does not follow the field and is (partially) transferred to other states. One might thus naively think that our observation is in a non-adiabatic regime. However, it appears that in a quantum mechanical non-adiabatic transition the phase of the system behaves erratically~\cite{Atkins:72}, while the phase is well-behaved in an adiabatic system. Since the phase of light in our microcavity varies smoothly, as evidenced by the appearance of interference fringes in the transient reflectivity spectrum \ref{fig:CmpMeasTheory}, we conclude that our experiments are in the adiabatic limit in quantum mechanics. Hence we propose to denote the observation that the frequency of light does not adiabatically follow the cavity resonance as \emph{not}-adiabatic to distinguish it from non-adiabatic behavior in quantum mechanics.

If we switch both GaAs and AlAs layers in our model, the transient reflectivity spectra are quite similar to the spectra for only GaAs switching (Fig.~\ref{fig:CmpMeasTheory}). Hence, spatially homogeneous or inhomogeneous switching is unimportant to the results. This confirms that our observations are in the adiabatic limit in quantum mechanics, since non-adiabatic dynamics is distinguished from adiabatic dynamics by the fact that the wavefunction changes symmetry, \emph{e.g.}, when spatial inhomogeneity is introduced~\cite{Notomi:06}.

Finally we briefly discuss several ways to apply and control our new frequency change mechanism.
i. Since the frequency changes away from the cavity resonance, frequency changes are greater than those achievable by adiabatic following. Currently, the trapped photons are red-shifted since the nature of the switching mechanism (excited carriers) has an increasing refractive index in time. Blue-shifted light is expected to occur if the refractive index decreases as can be realized with phonon-induced switching~\cite{Berstermann:09} or by electronic Kerr switching \cite{Ctistis:11}. ii. To increase the efficiency of the frequency change, it is desirable to provide faster relaxation times in the order of ps by optimized sample growth~\cite{Segschneider:97}, since $dE_{\rm{cav}}/dt$ is then even greater compared to $\Delta E_{\rm{cav}} / \tau_{\rm{cav}}$. iii. To maximize the ratio of bandwidth that is frequency shifted to the probe bandwidth, the probe's bandwidth could be significantly decreased. We expect the intracavity intensity, and therefore the transient reflectivity, to only depend on the spectral density of probe light near the instantaneous resonance wavelength, as long as the probe pulses remain much shorter than the cavity lifetime. Irrespective of the detailed free-carrier physics, we therefore expect similar results for pulses with a bandwidth down to 2 meV. iv. This scheme can easily applied in transmission mode. From our model, we find that a GaAs/AlAs microcavity in transmission mode predicts the same frequency shifts as in reflection. This is particularly attractive as the modulation depth varies with respect to 0 and not with respect to $100\%$. v. We find that the excess amplitude increases substantially by only doubling $Q$. A further increase of $Q$ saturates the excess amplitude. vi. Since the frequency-conversion occurs during the relaxation of the free carriers, $Q$ will always increase during the conversion process. Higher pump powers are therefore not expected to limit the switching efficiency. vii. Not-adiabatic cavity switching is suited to control frequency shifting of quantum states of light as the required strong pump field does not overlap with fragile quantum states. Hence our method opens new avenues in quantum optics and quantum information processing.

\section{Conclusions}
We have studied frequency-resolved femtosecond pump-probe reflectivity of a planar GaAs-AlAs microcavity. Between 5 and 40 ps after a pump pulse we observe a strong excess probe reflectivity. The frequency of light trapped in the cavity changes by up to 5 linewidths \emph{away} from the cavity resonance, and does not adiabatically follow the fast-changing cavity resonance. The frequency change is attributed to an accumulated phase change due to the time-dependent refractive index. An analytical model predicts dynamics in qualitative agreement with the experiments, and points to crucial parameters that control future applications.

\section{Acknowledgments}
We thank Bart Husken for help, and Ad Lagendijk, Klaus Boller, Pepijn Pinkse, Shanhui Fan for discussions. This work is part of the research program of FOM that is financially supported by NWO.
WLV thanks NWO-Vici, and Smartmix Memphis.

\bibliographystyle{osajnl2}

\begin{thebibliography}{99}
\bibitem{note:signal}
The measured spectrum $J(\omega)$ is the Fourier Transform of the time-dependent probe electric field ${\cal E}(t)$ that exits from the cavity:
\begin{eqnarray*}
J(\omega) &=& \pi r^2(\epsilon_0 c)^{-1}\left|\int_{-\infty}^{\infty}dt {\cal E}(t) e^{-i \omega
t}\right|^2,
\label{equa:signal}
\end{eqnarray*}
with $c$ the speed of light, $r$ the radius of the beam, and neglecting amplification and conversion factors.
Since the detection systems integration time of $150$ ns is much longer than a probe-sample interaction time the boundaries of the integral are taken to $\pm \infty$. Due to averaging of probe pulses over alternating pumped and unpumped events, the correct transient reflectivity is $2\times$ greater than raw data as in Figs. 2, 3; Fig. 4 has been corrected.
\end{thebibliography}

\end{document}